\documentclass[journal]{IEEEtran}
\usepackage{lipsum}

\ifCLASSINFOpdf
\else
\fi
\usepackage{amsmath}
\usepackage{amssymb}
\usepackage{graphicx}
\usepackage{svg}
\DeclareMathOperator{\sign}{sgn}
\newcommand{\multiline}[2]{\begin{tabular}{@{}c@{}}#1 \\ #2\end{tabular}}

\usepackage{comment}
\usepackage{tabulary}
\usepackage{multirow}

\usepackage[hidelinks]{hyperref}

\begin{document}
%
\title{Hidden Markov Models for Stock Market Prediction}
%
%
%

\author{Luigi~Catello, 
        Ludovica~Ruggiero,
        Lucia~Schiavone,
        and~Mario~Valentino
}

%
%

\markboth{Università degli Studi di Napoli Federico II - Identification and optimal control}{}
%



\maketitle

\begin{abstract}
\label{sec:abstract}
The stock market presents a challenging environment for accurately predicting future stock prices due to its intricate and ever-changing nature. However, the utilization of advanced methodologies can significantly enhance the precision of stock price predictions. One such method is Hidden Markov Models (HMMs). HMMs are statistical models that can be used to model the behavior of a partially observable system, making them suitable for modeling stock prices based on historical data. Accurate stock price predictions can help traders make better investment decisions, leading to increased profits.

In this article, we trained and tested a Hidden Markov Model for the purpose of predicting a stock closing price based on its opening price and the preceding day's prices. The model's performance has been evaluated using two indicators:  Mean Average Prediction Error (MAPE), which specifies the average accuracy of our model, and Directional Prediction Accuracy (DPA),  a newly introduced indicator that accounts for the number of fractional change predictions that are correct in sign.
\end{abstract}


%
\IEEEpeerreviewmaketitle

\section{Introduction}
\label{sec:intro}
\IEEEPARstart{P}{rediction} of the stock market, with its inherent volatility and potential for substantial financial gains, has long captivated the attention of institutional investors, hedge funds, and proprietary trading firms. These sophisticated market participants are driven by the desire to make accurate predictions about future price movements and trends in order to gain a competitive edge and maximize their investment returns.

\textit{Institutional investors}, such as pension funds and insurance companies, manage large pools of capital on behalf of their clients or beneficiaries. The primary objective of these investors is to generate consistent returns over the long term to fulfill their financial obligations. Accurate predictions in the stock market allow them to identify opportunities and mitigate risks, enhancing portfolio performance and meeting their fiduciary responsibilities \cite{institutional}.

\textit{Hedge funds}, on the other hand, are private investment partnerships that pool capital from accredited investors. Hedge fund managers seek to generate significant absolute returns, often irrespective of market conditions, by employing diverse investment strategies. Accurate predictions empower hedge funds to identify mispriced securities, exploit market inefficiencies, and construct profitable trading strategies, ultimately attracting investors and earning substantial profits \cite{hedgefunds}.

These sophisticated market participants employ a wide range of quantitative and qualitative methods to make predictions in the stock market. 

Algorithmic strategies rely on computer algorithms that systematically analyze vast amounts of market data, identify patterns, and generate trade signals. These algorithms are programmed to execute trades automatically, often leveraging complex mathematical models and statistical techniques. It is estimated that 50 percent of stock trading volume in the U.S. is currently being driven by algorithmic trading \cite{algo}.

While algorithmic trading offers significant advantages for institutional investors, retail investors should exercise caution when relying solely on algorithmic strategies. The success of algorithmic trading is often attributed to extensive research, robust infrastructure, and substantial resources that may not be readily available to individual retail investors. Moreover, market conditions can change rapidly, rendering algorithms ineffective or even exacerbating losses. Retail investors, lacking the same level of resources and risk management capabilities, face greater risks when solely depending on algorithmic trading.

In recent years, many strategies have been developed for algorithmic trading. Since Hidden Markov Models (HMMs) have emerged as powerful tools for the prediction of time series data, we expect them to give promising results in the context of the stock market prediction.

This paper is organized as follows: in Section \ref{sec:prev}, we briefly review various stock market prediction approaches; Section \ref{sec:approach} outlines our theoretical approach with relevant definitions and mathematical equations; in Section \ref{sec:impl} describes our technical implementation and solutions to encountered challenges; in Section \ref{sec:res} we present our results, compare them with other models, and introduce a novel evaluation metric; finally, in Section \ref{sec:conclusion} we summarize key findings and suggest directions for further developments.
\section{Previous Works}
\label{sec:prev}
Various approaches have been explored in the quest for reliable prediction models as accurate forecasting of stock prices and identification of trends are crucial for investors and financial institutions seeking informed decisions.

Before the emergence of artificial intelligence, probabilistic techniques formed the foundation of stock market forecasting.
These theories encompassed random walk models \cite{fama_random_1995}, \cite{lo_stock_2015}, correlation-based methods \cite{longin_correlation_1995}, \cite{ding_longmemory_1993}, scaling properties \cite{matteo_scaling_2003}, \cite{mantegna_scaling_1995}, stock market volatility and investor sentiment \cite{french_expected_1987}, \cite{lee_stock_2002}, probability distributions of stock price returns \cite{gabaix_theory_2003}, \cite{mantegna_applications_1999}, and other relevant approaches.

With the advancement of machine learning and the increasing computational power of computers, several techniques utilizing big data and artificial intelligence algorithms gained popularity.
Support Vector Machines (SVM) were quickly recognized as promising for time series prediction \cite{mukherjee_timeseries_1997}, and were first employed for financial forecasting in 2001 \cite{tay_svm_2001}.
Hybrid models combining ARIMA (Autoregressive Integrated Moving Average) and SVM were introduced by Pai et al. in 2005 \cite{pai_hybrid_2005}.
Neural networks, including specific types coupled with Genetic Algorithms, have been utilized to detect temporal patterns \cite{kim_genetic_2007}, \cite{kim_ga_2000}.
Bollen et al. scraped data from Twitter to forecast the stock market based on user mood in 2011 \cite{bollen_twitter_2011}.

In 2005, Hassan et al. employed Hidden Markov Models for time series prediction \cite{hassan_stock_2005}, building upon the concepts introduced in Rabiner's influential tutorial published in 1989 \cite{rabiner_hmm_1989}.
Their work demonstrated the effectiveness of HMMs, leading to their widespread adoption in subsequent studies \cite{zhang_high-order_2019}, \cite{zhou_datamining_2022}, \cite{hassan_combination_2009}.

In this project, we aimed to replicate and extend the findings reported by Gupta et al. in their 2012 study \cite{gupta_stock_2012}. We used Mathworks Statistics and Machine Learning Toolbox\texttrademark. Moreover, we decided to make our project open-sourced \cite{github_stock-market-prediction_2023}, hoping to foster collaboration and encourage the research community to build upon our work, replicate our experiments, and explore further enhancements to advance the field.  

\section{Approach}
\label{sec:approach}
\subsection{Markov Chains}
\label{subsec:mc}
A Markov chain is a stochastic model that represents a sequence of events or states. 
In our analysis, we will specifically focus on first-order Markov chains, which adhere to the Markov property. Let $S = \{S_1, S_2, \ldots, S_n\}$ be the set of all possible states, and $X = \{X_k | X_k \in S, k = 1, \dots, T\}$ be the states time series. The Markov property states that for any $k \geq 0$ and states $X_0, \dots, X_k$:
\begin{gather} 
P(X_{k+1} = S_j | X_k = S_i, X_{k-1}, \ldots, X_0 \\ \notag = P(X_{k+1} = S{j} | X_k = S_i)
\end{gather}
In other words, the probability of transitioning to a certain state $S_j$ at time step $k+1$ only depends on the current state $S_i$ at time step $k$ and not on any previous states. This property allows us to compute the probability distribution of the Markov chain at any future time step based exclusively on its current state.

Formally, a first order Markov chain is defined by the set of states $S$ and a transition probability matrix $A = [a_{ij}]$, where $a_{ij}$ represents the probability of transitioning from state $S_i$ to state $S_j$ in one step.
\begin{figure}[ht]
    \centering
    \includegraphics[width=150px]{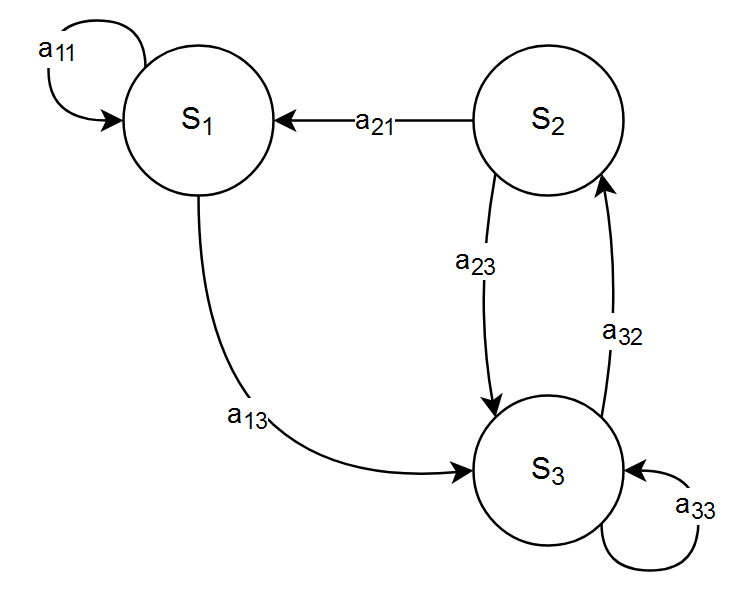}
    \caption{Example of a Markov Chain with three states.}
    \label{fig:markov-example}
\end{figure}

The elements of the transition matrix must satisfy the standard stochastic constraints:

\begin{itemize}
  \item $0 \leq a_{ij} \leq 1 \ \forall i, j$.
  \item $\sum_{j=1}^{n} a_{ij} = 1\  \forall i$.
\end{itemize}

Let $x_k \in \mathbb{R}^n$ be the vector containing the probabilities of being in each state at time $k$, the system evolves according to:
\begin{gather}
\label{eq:mc_evolution}
 x^T_{k+1} = x^T_kA
\end{gather}
\begin{sloppypar}
We refer to the initial distribution of probabilities as ${\pi=\{\pi_1,\pi_2\ldots \pi_n\}}$.
\end{sloppypar}

\subsection{Hidden Markov Models}
\label{subsec:hmm}

The model presented above implicitly assumes that each state corresponds to an observable (physical) event. Markov chains have proven to be valuable tools for modeling sequential data in various domains. However, many real-world scenarios involve underlying states that influence observations but are not directly observable. This limitation led to the development of Hidden Markov Models, which extend the basic Markov chain model by introducing hidden or unobservable states that affect the observed data.

The hidden state process of a HMM is a Markov chain, where each state generates an observation having a certain probability distribution that only depends on the state itself.

Let $O = \{O_k | O_k \in \mathcal{O}, k = 1, \dots, T\}$ be the observed sequence, where $\mathcal{O}$ is the set of possible observations. The hidden states are assumed to evolve according to Equation \ref{eq:mc_evolution}.

The probability of symbol $o\in \mathcal{O}$ being emitted by state $S_i$ is described by the emission probability function $ b_i : o  \mapsto b_i(o) \in [0, 1], \forall i$.

Both in the current and following sections, we refer to the emission probability matrix $B=\{b_{ij} = b_i(o_j)\}$.

\begin{figure}[ht]
    \centering
    \includegraphics[width=150px]{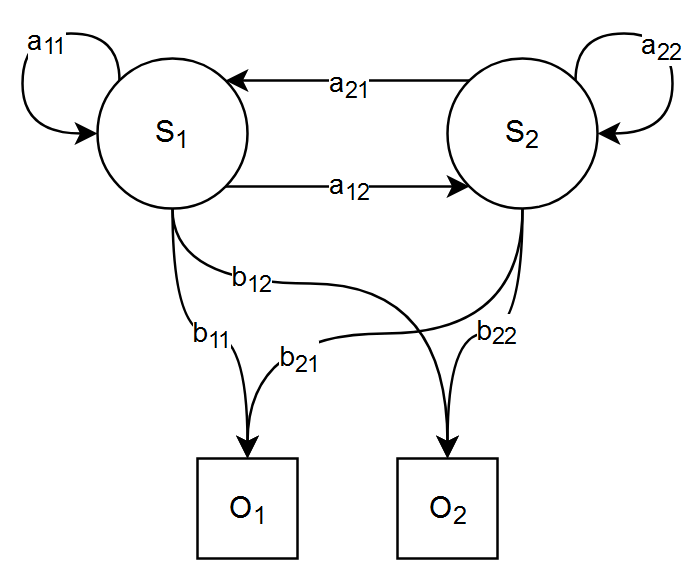}
    \caption{Hidden Markov Chain Example with two states and two possible outputs.}
    \label{fig:hmm_example}
\end{figure}

\subsection{Applications of HMMs: Common Problems and Solutions} 
\label{subsec:hmm-applications}

The most common problems analyzed using HMMs include the evaluation problem, decoding problem, and parameter estimation problem. These problems and their corresponding solutions have been extensively studied and documented in the literature \cite{rabiner_hmm_1989}.
In the following paragraphs, we provide an overview of these problems and their solutions. 

\subsubsection{Evaluation Problem}
\label{subsub: evaluation problem}
The evaluation problem in HMMs involves computing the probability of a known observation sequence, given the model. Specifically, given an HMM with the parameters $\pi$, $A$, and $B$, and an observed sequence $O$, we want to compute $P(O|\pi, A, B)$.

The procedure used to solve this problem involves computing the forward variable, denoted as $\alpha_k(i)$, which represents the probability of being in  state $S_i$ at time $k$ and observing the sequence up to time $k$. It is computed recursively for each time step $k$ and state $S_i$.

\begin{sloppypar}
At the initialization step, the forward variable for the initial time step ${k=1}$ is calculated as: ${\alpha_1(i) = \pi_i \cdot b_{i} (O_1),} \,\,{\forall i}$,
where $\pi_i$ is the initial probability of being in state $S_i$ and  $b_{i}(O_1)$ is the emission probability for the first observation $O_1$.
\end{sloppypar}

For each subsequent time step $k > 1$, the forward variable $\alpha_k(i)$ is computed as the sum of the probabilities of all possible paths that could 
have generated the observations up to time $k$ and reached state $S_i$. This is expressed by the following formula:
\begin{equation}
    \alpha_k(i) = \left(\sum_{j=1}^{n} \alpha_{k-1}(j) \cdot a_{ji}\right) \cdot b_{i}(O_k), \quad \forall i
\end{equation}

By recursively calculating the forward variables for each time step, the algorithm evaluates the probability of observing the entire sequence, given the HMM parameters.

Finally, to obtain the overall probability of the observed sequence, the forward procedure computes the sum of the forward 
variables at the final time step $T$:

\begin{equation}
    P(O|\pi, A, B) = \sum_{i=1}^{n} \alpha_T(i)
\end{equation}

\subsubsection{Decoding Problem}
\label{subsub: decoding problem}
The decoding problem involves determining the most likely sequence of hidden states given an observation sequence and the model. Given an HMM with the parameters $\pi$, $A$, and $B$, and an observed sequence $O$, we want to find the hidden state sequence $X$ that maximizes $P(X|O, \pi, A, B)$.

The solution to the decoding problem is commonly addressed using the Viterbi algorithm.

The algorithm works by iteratively calculating the most likely path to each state at each time step, considering both the current observation and the previous states probabilities. At each time step, it computes the probability of being in each state and tracks the most probable path leading to that state.

\subsubsection{Parameter Estimation Problem}
\label{subsub: parameter estimation}
The parameter estimation problem in HMMs involves adjusting the model parameters to maximize the probability of an observed sequence. Given an observation sequence $O$, we want to estimate the optimal values for the parameters $\pi$, $A$, and $B$ that maximize $P(O|\pi, A, B)$.

The solution to the this problem is typically addressed using the forward-backward algorithm, also known as the Baum-Welch algorithm.

The Baum-Welch algorithm is a specific implementation of the Expectation–Maximization (EM) algorithm, tailored for HMMs. It iteratively performs three main steps:  forward, backward, and update step. In the forward step, the algorithm computes the probability of being in a specific state at each time step, given the observed sequence up to that point.
This is the evaluation problem explained before.

In the backward step, the algorithm computes the probability of observing the remaining part of the sequence from a given state at each time step. It complements the forward procedure by computing the \emph{backward variable}, denoted as $\beta_k(i)$, which represents the probability of being in state $S_i$ at time $k$ and observing the remaining sequence from time $k+1$ to the end.
The backward variable is computed recursively for each time step $k$ and state $S_i$. At the initialization step, the backward variable for the final time step is set as:   $\beta_T(i) = 1, \, \forall i$.

For each time step $k < T$, the backward variable $\beta_k(i)$ is computed as the sum of the probabilities of all possible paths that can be generated from state $S_i$ at time $k$ to the end of the sequence. This is expressed by the following formula:
\begin{equation}
    \beta_k(i) = \sum_{j=1}^{n} \beta_{k+1}(j) \cdot a_{ij} \cdot b_{j}(O_{k+1})
\end{equation}

After computing the forward and backward variables, they are used to refine the parameters of the HMM based on the observed sequence.
In fact, let $\xi_k(i,j)$ be the probability of being in state $S_i$ at time $k$ and state $S_j$ at time $k+1$, given the model and the observation sequence. It can be computed as:
\begin{align}
\xi_k(i,j) &= \frac{P(X_k=i,X_{k+1}=j,O|\pi, A, B)}{P(O|\pi, A, B)}  \\
           &=\frac{\alpha_k(i)\cdot a_{ij} \cdot b_{j}(O_{k+1})\cdot \beta_{k+1}(j)}{\sum_{r=1}^{n}{\sum_{h=1}^{n}{\alpha_k(r)\cdot a_{rh}\cdot b_{h}(O_{k+1})\cdot \beta_{k+1}(h)}}} \notag
\end{align}
Let's also  define $\gamma_k(i)$ as the probability of being in state $S_i$ at time $k$, given the observation sequence and the model:
\begin{align}
\gamma_k(i) &= P(X_k=i|O,\pi, A, B) \notag \\
           &=\frac{\alpha_k(i)\cdot\beta_{k}(i)}{\sum_{j=1}^{n}{\alpha_k(j)\cdot\beta_{k}(j)}}
\end{align}
We can relate those quantities:
\begin{equation}
    \gamma_k(i)=\sum_{j=1}^{n}{\xi_k(i,j)}
\end{equation}
Moreover, $\sum_{k=1}^{T-1}{\gamma_k(i)}$ represents the expected number of transitions from state $S_i$, and $\sum_{k=1}^{T-1}{\xi_k(i,j)}$ represents the expected number of transitions from state $S_i$ to state $S_j$.

The update step involves estimating the new values for the initial state probabilities, transition probabilities, and emission probabilities.
For the initial state probabilities, the updated values are calculated as the normalized forward-backward probabilities at the initial time step:
\begin{equation}
\pi_i = \gamma_1(i)
\end{equation}

The transition probabilities are updated based on the expected number of transitions between states, normalized over the expected number of transitions from each state
\begin{equation}
    a_{ij} = \frac{\sum_{k=1}^{T-1}{\xi_k(i,j)}}{\sum_{k=1}^{T-1}{\gamma_k(i)}}
\end{equation}

The estimate for the emission probability $b_{j}(o)$ --- the probability of observing symbol $o$ from state $S_j$ --- is calculated by evaluating the expected number of times in which $o$ is emitted from state $j$. This value is then normalized by the expected number of transitions from state $S_j$:
\begin{equation}
    b_{j}(o) = \frac{\sum_{k=1}^{T} \Big(\gamma_k(j) \cdot \delta(O_k,o)\Big)}{\sum_{k=1}^{T} \gamma_k(j)}
\end{equation}
where $\delta(O_k,o)$ is the Kronecker delta function that evaluates to 1 when $O_k$ is equal to the observed symbol $o$, and 0 otherwise.

This process iteratively refines the model parameters until convergence, where they stabilize, meaning that the likelihood of the sequence is maximized.

Choosing the initial conditions wisely is crucial as it can significantly impact the estimation of the HMM parameters. By selecting appropriate initial conditions, the Baum-Welch algorithm is more likely to converge to some parameters that provide accurate modeling of the underlying dynamics.

\subsection{Gaussian Mixture Models}
\label{subsec:gmm}

Gaussian Mixture Models (GMMs) are powerful probabilistic models used to represent complex probability distributions. A GMM is a weighted sum of multiple Gaussian distributions, where each component represents a subpopulation within the data.

When a single observation is composed by multiple data (multivariate data), GMMs are employed as multivariate mixture models. Each Gaussian component in a GMM represents a multivariate distribution with its own mean vector and covariance matrix. The probability density function of a multivariate GMM is given by:
\begin{equation}
p(y) = \sum_{i=1}^{K} c_i\, \mathcal{N}(y|\mu_i, \Sigma_i)
\end{equation}
where $K$ is the number of Gaussian components, $c_i$ represents the weight associated with the $i$-th component, and $\mathcal{N}(y|\mu_i, \Sigma_i)$ denotes the multivariate Gaussian distribution with mean vector $\mu_i$ and covariance matrix $\Sigma_i$.

The weights $c_i$ satisfy the constraint $\sum_{i=1}^{K} c_i = 1$, ensuring that the probabilities sum up to one. These weights control the contribution of each Gaussian component to the overall distribution.

GMMs are particularly useful when modeling complex multivariate data that exhibit multiple modes or clusters because they can capture the inherent structure and variability present in the data.

In the context of Hidden Markov Models, GMMs are employed to initialize the emission probabilities of the hidden states.
The GMM is fitted on the training dataset: the training algorithm estimates the parameters of each Gaussian component, i.e. the mean vectors and covariance matrices.
These parameters are used to initialize the emission probabilities of the corresponding hidden states.

\section{Technical implementation}
\label{sec:impl}

In our study, three specific functions provided by the Statistics and Machine Learning Toolbox\texttrademark\  in MATLAB were utilised for the training and testing of the HMM: \verb|hmmtrain|, \verb|hmmdecode| and \verb|fitgmdist|.

The model was trained and tested on the historical daily prices of Apple, IBM and Dell stocks that are publicly available on the web.
Each observation in our dataset comprises three distinct values representing the daily fractional change, fractional high, and fractional low prices.
\begin{gather*}
    O_k = \left( \frac{Close - Open}{Open},\frac{High-Open}{Open},\frac{Open - Low}{Open} \right)
\end{gather*}    
\begin{gather}
    O_k:=\big(fracChange,fracHigh,fracLow\big) 
\end{gather}

The array $O_k$  is a three-dimensional array consisting of real values.
Since the probability of guessing any real value is mathematically zero, it was necessary to discretize the observations.
The number of points used for the discretization is specified in Table \ref{tab:discretization}.

\begin{table}[htb]
 \centering
 \caption{Number of discrete points for each dimension}
 \label{tab:discretization}
 \begin{tabular}{|c|c|}
   \hline
   \textbf{Variable} & \textbf{Number of points} \\
   \hline
   fracChange & $n_{\text{fC}} = 50$ \\
   \hline
   fracHigh & $n_\text{fH} = 10$\\
   \hline
   fracLow & $n_\text{fL} = 10$ \\
   \hline
 \end{tabular}
\end{table}

We used dynamic edges for the discretization: at each train, the maximum and minimum values of \textit{fracChange}, \textit{fracHigh}, and \textit{fracLow} are computed.
We then generated three linearly spaced vectors for the edges, using as the lowest values the $\min$, and as the highest values the $\max$ values. Finally, we assigned to each discrete value its corresponding index in the edges array.

Furthermore, compatibility with our monovariate Hidden Markov Model required mapping the three-dimensional space to a one-dimensional space. 
The mapping process was accomplished by enumerating the points within the discrete three-dimensional set, as described in Equation \ref{eq:mapping_eq}.
The inverse mapping was achieved using Equations \ref{eq:inverse_mapping_eq}: 
\begin{gather}
\label{eq:mapping_eq}
    n = (z-1)\cdot(x_{max}\cdot y_{max}) + (y-1)\cdot x_{max} + x
\end{gather}

\begin{align}
\label{eq:inverse_mapping_eq}
     & z = \left\lfloor \frac{n-1}{x_{max}\cdot y_{max}}\right\rfloor +1    \notag \\    
     & y = \left\lfloor \frac{n-1 - (z-1)\cdot x_{max} \cdot y_{max}}{x_{max}} \right\rfloor + 1\\
     & x = (n-1) \bmod (x_{max} )+ 1\notag
\end{align}

\subsection{Training}
\label{subsub:train}
During the initial trains we assumed four underlying hidden states, where each state generates outputs represented by a GMM with four components. We then re-trained our model varying the values. The specific parameters used for each train can be found on our GitHub repository \cite{github_stock-market-prediction_2023}.
The parameters of these GMMs were estimated using the MATLAB function \verb|fitgmdist|, which optimizes the model using the Expectation-Maximization (EM) algorithm. The initial parameters of the GMM are obtained through \textit{k-means} clustering. The resulting probability density function served as the initial estimate for the emission matrix. Moreover, we assigned a uniform distribution of probabilities as the initial estimate for the transition matrix.

The training data for the HMM is constructed by using a rolling window approach. In this approach, each observation sequence spans a fixed duration of 10 days. We refer to this duration as \textit{latency}. The window is shifted incrementally along the training period: the first sequence captures observations from the initial time point, while each following sequence incorporates new observations by sliding the window by one day. The dataset is then provided as input to the MATLAB function \verb|hmmtrain|. This function utilizes the Baum-Welch algorithm to estimate the transition and emissions matrices. These estimations are initialized with the initial guesses discussed previously.

\subsection{Prediction}
\label{subsub:pred}
Following the training phase, we proceeded to test our model by predicting the stock daily close prices for different time frames. For each day $d$ within the target period, the prediction process involved the following steps:
\begin{enumerate}
    \item We considered the last $latency - 1 = 9$ observations available.
    These observations represent the preceding 9 days.
    \item Next, we appended each possible output for the current day $d$, creating a 10-day sequence. This sequence now encompasses the 9 historical observations and one potential observation for the next day. There are $n_\text{fc}\cdot n_\text{fH} \cdot n_\text{fL}$ possibilities for the current day.
    \item We computed the probability for each sequence to be generated from our trained model. Finally, we selected the observation with the highest emission probability as the observation for the next day.
\end{enumerate}

In certain cases, there may arise situations where the probability of emitting the historical observations, along with any hypothetical observation, becomes 0 or very close to 0. This can occur due to numerical errors or limitations inherent the model. However, we have found that incorporating a \textit{dynamic window} can enhance the performance of our model.

To address the issue, we have modified the prediction algorithm as follows: if the highest probability obtained is $0$, we repeat the prediction algorithm while gradually reducing the latency by one day. By reducing the latency, we aim to find a viable solution where the emitted probabilities are non-zero. This process of reducing latency is repeated iteratively until a solution is found, while ensuring that the historical sequence remains sufficiently long. In our case, we have set a minimum requirement of four days for the historical sequence.

\section{Results}   \label{sec:res}

\begin{figure*}
    \centering
    \includegraphics[width=0.75\textwidth]{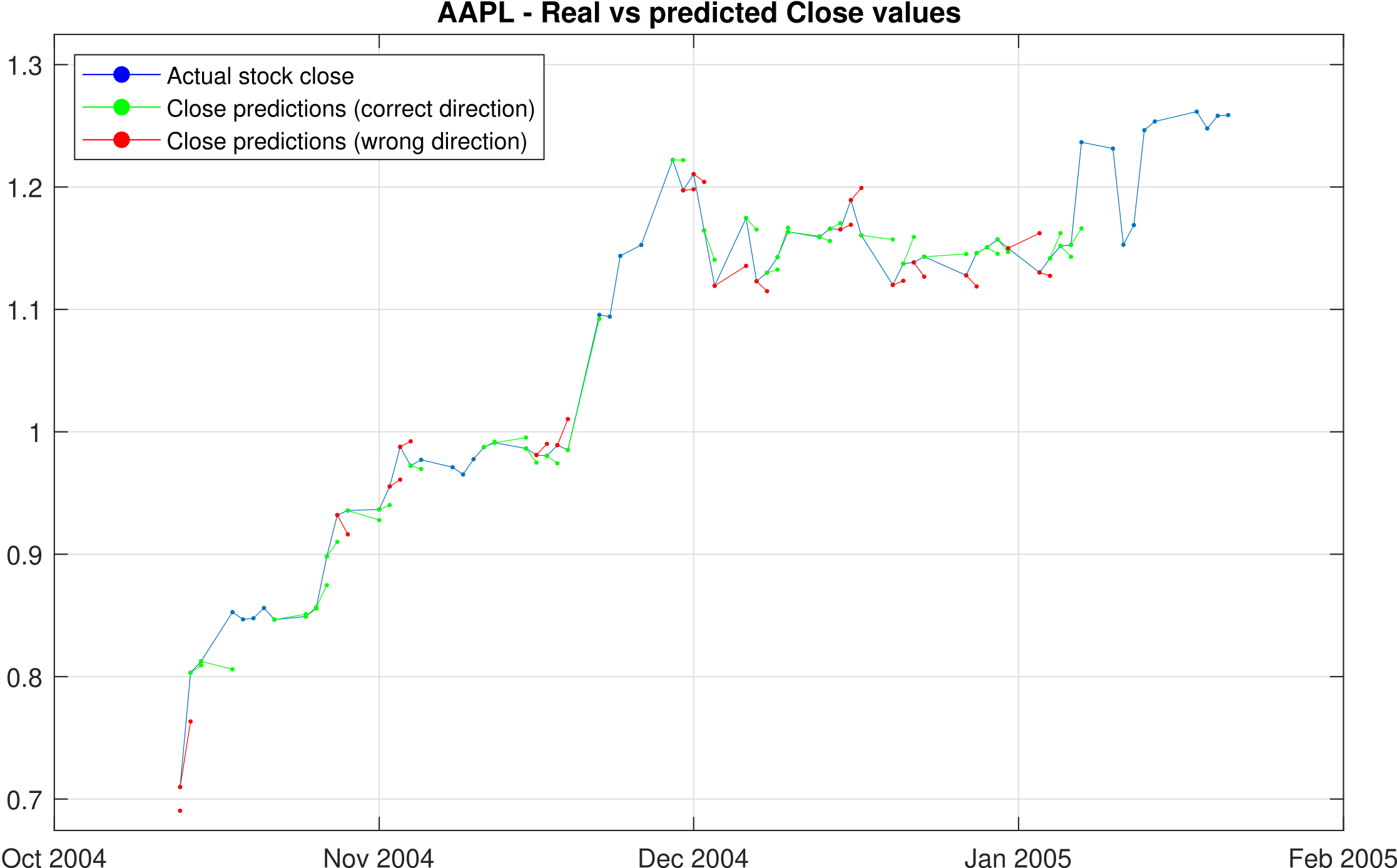}
    \caption{Actual vs Forecasted close prices for AAPL.}
    \label{fig:aapl_close}
\end{figure*}

We used multiple metrics to evaluate the performance of our models.
Mean Absolute Percentage Error (MAPE) is calculated between the actual and predicted stock closing prices.
\begin{equation} \label{eq:mape}
    \text{MAPE} = \frac{1}{n_p} \sum_{i = 1}^{n_p} \frac{\left|p_i - c_i \right|}{\left| c_i \right|} \cdot 100\%
\end{equation}
On day $i$, $p_i$ is the predicted stock closing price, $c_i$ is the actual stock closing price, and $n_p$ is the total number of predictions.
Table \ref{tab:mape} shows our results on two different stocks, compared to the results from other papers \cite{gupta_stock_2012}, \cite{hassan_combination_2009}.

\begin{table}[ht]
\renewcommand{\arraystretch}{1.2}
\caption{Comparison of Mean Absolute Percentage Error}
\label{tab:mape}
\centering
\begin{tabulary}{\linewidth}{|C|C|C|C|C|}
\hline
\textbf{Stock Name} & \textbf{Our HMM} & \textbf{HMM-MAP \cite{gupta_stock_2012}} & \textbf{HMM + Fuzzy Model \cite{hassan_combination_2009}} & \textbf{ARIMA / ANN}\\
\hline
Apple Inc. & 1.50 & 1.51 & 1.77 & 1.80\\ 
\hline
IBM Corp. & 0.68 & 0.61 & 0.78 & 0.97\\
\hline
\end{tabulary}
\end{table}

Predicting the movement of stock prices for a single day is a challenging task in itself.
Achieving accuracy in predicting multiple consecutive days becomes even more difficult and often approaches the realm of impossibility.
Recognizing the potential utility of predicting whether the stock value will increase or decrease during the day, we introduced a novel evaluation metric called Directional Prediction Accuracy (DPA). 
DPA measures the percentage of correct directional predictions, providing valuable information about the accuracy of our model.
\begin{equation} \label{eq:dpa}
    \text{DPA} = \frac{1}{n_p} \sum_{i=1}^{n_p} \delta\big(\sign(p_i - s_i), \sign(c_i - s_i)\big) \cdot 100\%
\end{equation}
In Equation \ref{eq:dpa}, $n_p$ is the total number of predictions, $\delta$ is the Kronecker delta function, $p_i$ is the predicted stock closing price, $c_i$ the actual closing price, and $s_i$ is the stock opening price.
Table \ref{tab:dpa} demonstrates that MAPE and DPA are not equivalent measures as they are in general not correlated.

\begin{figure*}
    \centering
    \includegraphics[width=0.75\textwidth]{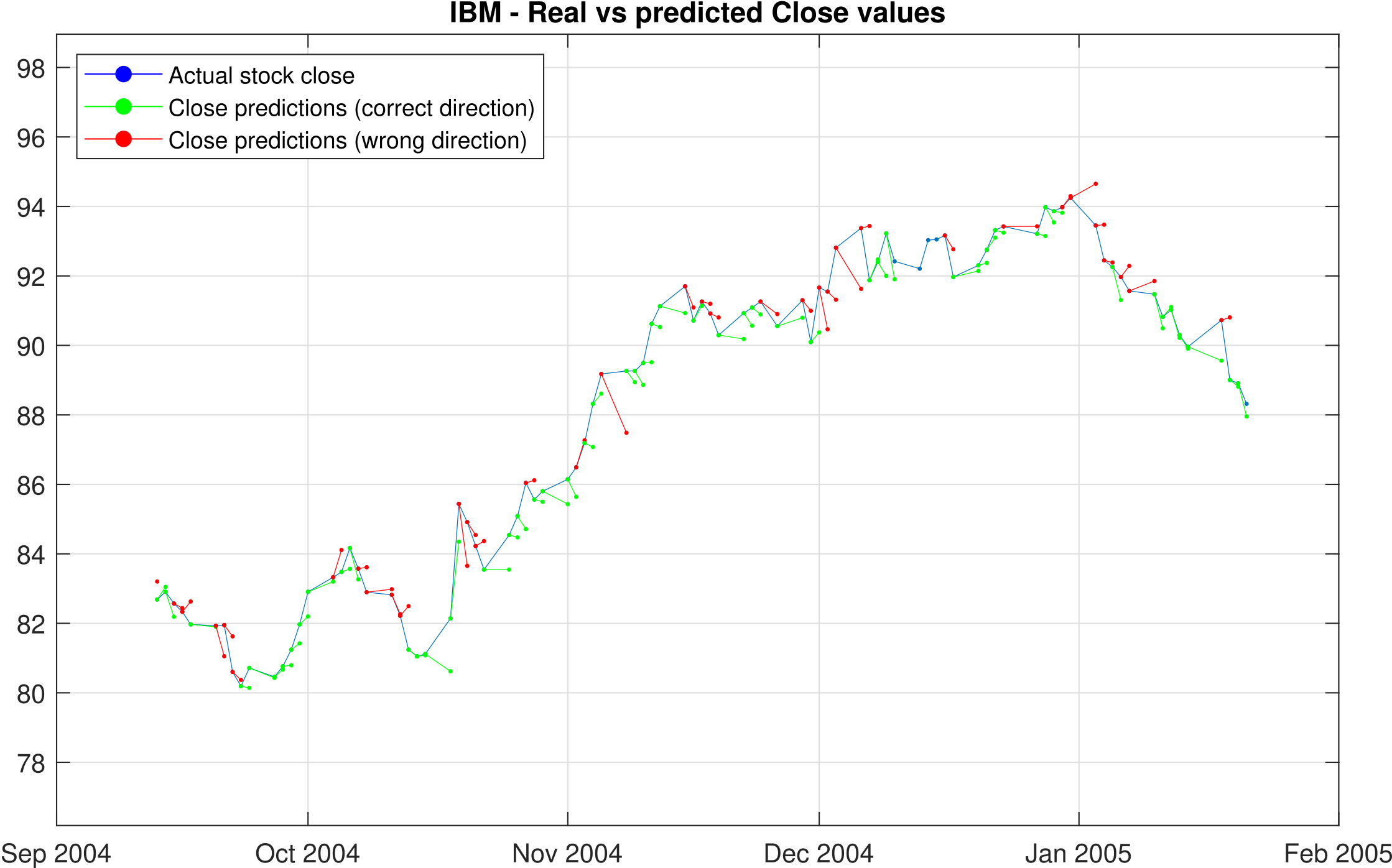}
    \caption{Actual vs Forecasted close prices for IBM } 
    \label{fig:IBM_close}
\end{figure*}

\begin{table}[!h]
\caption{Directional Prediction Accuracy}
\label{tab:dpa}
\centering
\begin{tabulary}{\linewidth}{|C|C|C|C|C|}
\hline
\textbf{Stock Name} & \textbf{MAPE} & \textbf{DPA} & \textbf{Training} & \textbf{Testing}\\
\hline
{Apple Inc.}&1.50\% &52.11\%&\multiline{2003-02-10}{2004-09-10}& \multiline{2004-09-13}{2005-01-21}\\ \cline{2-5}
\hfill\footnotemark[1]&1.73\%&63.27\%&\multiline{2003-02-10}{2004-09-10}&\multiline{2004-10-13}{2005-01-21}\\ \cline{2-5} 
&1.05\%&53.06\%&\multiline{2021-01-04}{2022-01-03}&\multiline{2023-01-03}{2023-06-30}\\ \cline{2-5} 
&1.01\%&40.57\%&\multiline{2017-01-03}{2019-01-03}&\multiline{2023-01-03}{2023-06-30}\\ \hline 

{IBM Corp.}&0.77\%&54.55\%&\multiline{2003-02-10}{2004-09-10}&\multiline{2004-10-13}{2005-01-21}\\ \cline{2-5} 
&0.82\%&57.58\%&\multiline{2003-02-10}{2004-09-10}&\multiline{2004-10-13}{2005-01-21}\\ \cline{2-5} 
&0.68\%&60.23\%&\multiline{2003-02-10}{2004-09-10}&\multiline{2004-09-13}{2005-01-21}\\ \cline{2-5} 
\hfill\footnotemark[2]{}&0.88\%&52.73\%&\multiline{2021-01-04}{2023-01-03}&\multiline{2023-01-04}{2023-07-11}\\ \hline

Dell Inc. &1.45\%&60.32\%&\multiline{2021-01-04}{2022-01-03}&\multiline{2023-01-03}{2023-07-11}\\ \hline
\end{tabulary}
\end{table}
\footnotetext[1]{Figure \ref{fig:aapl_close} shows predicted and actual close values for this train. Green lines indicate predictions that guessed right the sign of the fractional change with respect to the opening price. Therefore, 63.27\% of the predictions are correct in sign. The high MAPE reflects the accuracy of the predictions that often deviate from the actual close value, although being correct in sign.}
\footnotetext[2]{The results of this train are shown in Figure \ref{fig:IBM_close}: since the MAPE is low, predicted close values are, on average, more accurate.}

\begin{figure}[!t]
    \centering
    \includegraphics[width=\linewidth]{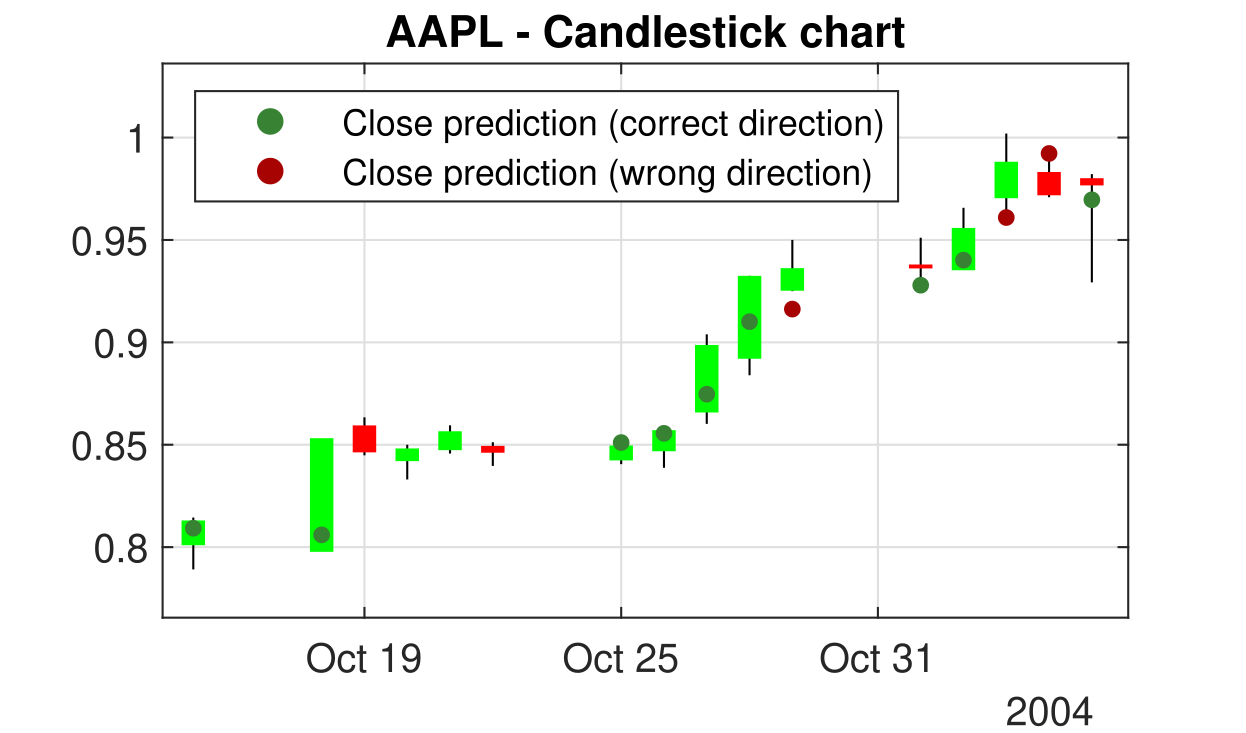}
    \caption{Candlestick chart along with predicted close values for AAPL}
    \label{fig:aapl_candle}
\end{figure}

Figure \ref{fig:aapl_candle} remarks the difference between the two indicators: the candlestick chart shows the actual prices for AAPL during the testing period 2004-10-13 to 2005-01-21. The dots show the predicted close values for the corresponding day. When the dots are green, they contribute to increasing the DPA. They only improve the MAPE when they get closer to the actual close value.


\section{Conclusions and Further Developments}
\label{sec:conclusion}
In this study, we have undertaken a comprehensive implementation of the Stock Prediction Hidden Markov Model originally proposed by Gupta et al. \cite{gupta_stock_2012}. Building upon their foundational work, we aimed to assess the model's performance on diverse datasets and explore its effectiveness in predicting stock market movements. By carefully replicating and extending their approach, we have conducted a rigorous evaluation, comparing the results with benchmark models to gain valuable insights into the model capabilities and limitations.

Our model was trained on a time period of one to two years and used to make predictions on a different time span, demonstrating its flexibility and re-usability. The complete source code, along with pre-trained models and a summary of their characteristics, is available on GitHub \cite{github_stock-market-prediction_2023}.

Our implementation underwent rigorous testing on various stocks, and we compared the results with those obtained from HMM-MAP, HMM-Fuzzy, ARIMA, and ANN models. We conducted thorough exploration of different hyperparameters, measuring their impact in search of optimal values. The outcomes demonstrated a significantly lower Mean Absolute Percentage Error compared to HMM-Fuzzy, ARIMA, and ANN models when trained on the same years as those in \cite{gupta_stock_2012} and \cite{hassan_combination_2009}.
Furthermore, to gain additional insights into the efficiency of our model, we developed a novel evaluation metric named Directional Prediction Accuracy (DPA). The DPA allowed us to assess the accuracy of our predictions in capturing stock price movements, providing valuable information for model performance assessment.

For further improvement, we propose fine-tuning the Emission and Transmission matrices on the latest time window each day before making predictions. This approach involves training the main model on historical data, then updating the matrices with data from a more recent period just before making each prediction. For example, we could train the main model from 2021-01-01 to 2023-01-01 and fine-tune the matrices using data from 2022-06-01 to 2023-06-01 before making a prediction for 2023-06-02. This process could potentially capture more recent market trends and improve the accuracy of predictions.

In conclusion, our implementation of the Stock Prediction HMM exhibited strong predictive capabilities, outperforming other benchmark models. The proposed fine-tuning approach holds promise for enhancing future predictions and warrants further investigation to capitalize on recent market dynamics. Our work contributes to the growing field of stock market prediction and provides valuable insights for traders and financial analysts.

%
%
%
%


%
\hfill July 2023
\ifCLASSOPTIONcaptionsoff
  \newpage
\fi



\bibliographystyle{IEEEtran}
\bibliography{bibtex/bib/paper}
\end{document}